# PrePrint 05Mar08
# Investigation of the Vertical Movement of an Isothermal Line at the Density Maximum in $H_2O$ and $D_2O$

William R. Gorman[a)], Gregory J. Parks[b)], and James D. Brownridge[c)]
State University of New York at Binghamton
Department of Physics, Applied Physics, and Astronomy
Binghamton, NY 13902

We studied the cooling of a column of water, primarily in a freezer, and analyzed the development and movement of an isothermal line at $\sim 4°C$ in $H_2O$ and $\sim 11°C$ in $D_2O$. Our experiments show that the vertical velocity of the symmetrical isothermal line moving up the column of water is inversely proportional to the diameter of the column of water. It has a measured maximum velocity of 1.4 ± 0.1 cm/min when the column diameter is 22 mm and decreases to 0.4 ± 0.1 cm/min when the diameter is increased to 125 mm. The measurement of the velocity becomes increasingly difficult to obtain when the column diameter is less than 22 mm because of the lack of complete development of the isothermal line. The data and discussion presented in this paper raise serious questions to the claim of new phase transitions in water made by S. Esposito, et. al.[1]

## INTRODUCTION

At first glance the water molecule ($H_2O$) appears to be a simple compound but with the x-ray technology developed in 1930 it was realized how complicated water is. Water is such a complex compound that to date it has sixty-four known anomalies listed in the literature.[2-11] Its density maximum of 0.99995 g/mL at $3.984°C$ [3] is one of the unique anomalies which can be indirectly observed when water is cooled from above $4°C$ to below $4°C$. In an attempt to better understand just how water behaves when it is cooled through its density maximum, we have investigated the vertical movement of thermal fronts in columns of water as they are cooled from about $20°C$ to near freezing. Upon cooling water in a column, a cold front is spontaneously produced when water at the bottom of the column transitions through its maximum density at $\sim 4°C$. The cold front is essentially an isotherm that moves up the column of water where the temperature above the isotherm is greater than $4°C$ and the temperature below the isotherm is less than $4°C$. Although there are many references in the literature to $\sim 4°C$ isotherms in cooling and warming water, we do not find any studies dealing with direct experimental measurement of the velocity of the $\sim 4°C$ isotherm moving up a column of confined still water[12-14]. By still water we mean there is no external turbulence induced in the water before the onset of cooling. In this study we observe the development and measure the velocity of the $\sim 4°C$ isotherm which occurs in still water. We do this in an attempt to better understand how still water, confined to a container, cools.

## EXPERIMENTAL PROCEDURE

The effect being described as a cold front moving up through the water was first noticed when de-ionized (DI) $H_2O$ was cooled by means of an ice bath. The container, which had been varied in length and diameter, had 7 type T thermocouples evenly spaced down the length of the container such that the tip of the thermocouple would extend to the containers center, Fig. 1a. The two containers used were cylinder

---

[c)] jdbjdb@binghamton.edu

in shape and had the following dimensions: 1) 15 cm long, 6 cm diameter; and 2) 128 cm long, 7.5 cm diameter.

The containers were filled with DI H$_2$O and allowed to sit undisturbed at room temperature for 10 minutes. During that time a similar empty container was placed vertically in the ice bath to make a pocket for the filled one. After the 10 minutes, the empty container was replaced by the filled one. It was done in such a manner as to not induce turbulence in the container because the cold front effect is best observed in still water. There was an 8$^{th}$ thermocouple placed in an external ice bath to correct for electronic drift.

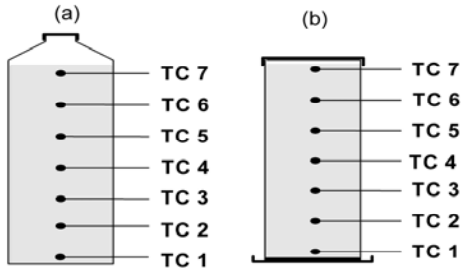

**FIG. 1**. Schematic diagram of the water containers used to measure the vertical velocity of isothermal lines; **(a)** container for ice bath cooling and **(b)** container for freezer cooling. TC 1-7 are seven thermocouples positioned axially in the center of each container with TC 1 at the bottom.

The claim that the cold front is best observed in still water was also tested. This test was done with a 5 cm wide (9 cm tall) container. A vertical rod, with a paddle at its tip, had been placed in the container such that the paddle was near the bottom of the container. This paddle rotated at a rate of 7 rpm which stirred the water. A thermocouple was placed at 6.5 cm from the bottom of the container. The cooling of the stirred water takes the form of Newton's Law of Cooling,

$$\Delta T = \Delta T_o \, e^{-t/\tau} \qquad (1)$$

where $\Delta T_o$ is the initial temperature difference at t=0 and $\tau$ is the time constant. This says that the change in temperature between two objects, i.e. a container of water placed in a cold environment, will be in the form of an exponential.[15]

For increased stability and reproducibility we relied primarily on freezer cooling. This gave us a more consistent cooling rate and we did not have to worry about induced turbulence just prior to cooling. The reproducible cooling rate of the freezer is expressed by

$$T = A_1 \, e^{(-t/\tau)} + T_o \qquad (2)$$

with $T$ being the temperature ($°C$), $A_1$ the amplitude of $39 \pm 1°C$, $t$ is the time in seconds, $\tau$ the time constant of $1576 \pm 44$ seconds, and $T_o$ the minimum temperature of $-13 \pm 1°C$. This equation (2) was determined by an analytical fit of experimental data from the cooling of the freezer. For the freezer experiments the water containers were glass and plastic cylinders of similar height but of varying inside diameter. We ran several different tests with this setup which are as follows: 1) with a thermocouple centered at 0.5 cm from the top and 0.5 cm from the bottom, in a 20 cm tall cylinder, we varied the cylinder diameter to see if that affected the development of the cold front; 2) with 3 thermocouples vertically centered, starting 1 cm from the top and spaced 3 cm apart, we varied the diameter and measured the speed of the cold front; and 3) we placed 5 thermocouples horizontally, spaced 0.8 cm apart at 4 cm from the top, to determine the shape of the cold front. The placement of the horizontal thermocouples was chosen to be 4 cm from the top because previous experiments suggested that the front becomes most defined at ~20% from the top. An illustration of the cylinders used can be seen in Fig. 1b. These containers were filled with room temperature DI H$_2$O and placed in the freezer that was off and also at room temperature. A thermocouple connected to a small piece of copper (~0.5 cm$^3$) was suspended in the freezer to monitor its air temperature and another was placed in an ice bath to correct for any electrical drift. The cylinders inside diameter varied from 4 mm to 125 mm. All figures of temperature versus time have been smoothed by a 15 point adjacent averaging function.

## RESULTS

For the ice bath test, the results for the 60 mm diameter cylinder are shown in Fig. 2. The other cylinder produced a similar graph. It clearly shows that there is a sudden drop in temperature as time increases which corresponds to the cold front passing each thermocouple on its way to the top of the cylinder. The DI H$_2$O was heated in this run to show that the effect is not initial temperature dependent.

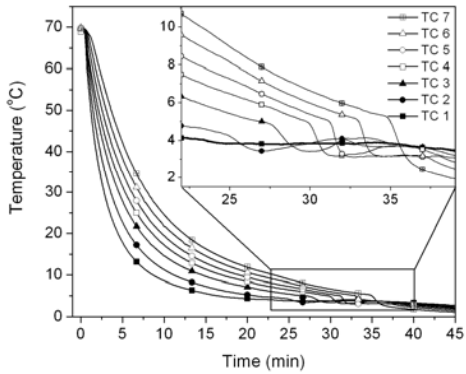

**FIG. 2.** Temperature versus time of warm water cooled in an ice bath in a container with a 60 mm diameter and 15 cm length. **Inset:** Close up of the movement of the cold front from that same run. TC 1 is at ~0.3 cm from the bottom of the container and TC 7 is at ~0.3 cm from the top of the container.

With the thermocouples being ~3 cm apart it is very obvious that there is a certain amount of time between the sudden drops which corresponds to the speed of the cold front.

To confirm that the water must be still in order for the cold front to develop we compared still water with stirred water. The stirred water follows Newton's Law of Cooling, (1), with a time constant of 415 ± 24 seconds. The differences in the curves are shown in Fig. 3 which clearly shows that the still $H_2O$ does not directly follow (1).

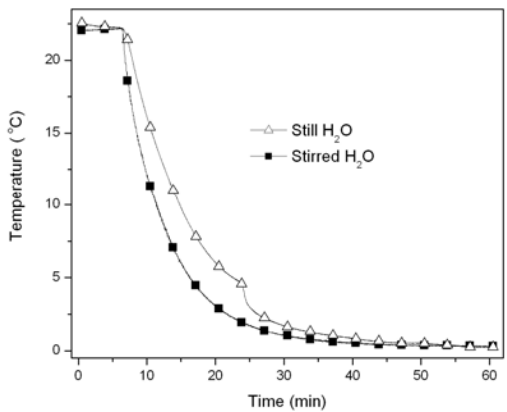

**FIG. 3.** Temperature versus time showing the difference between stirred and non-stirred water. The thermocouple is at the same location for each curve and the error bars are smaller than the symbols.

For the freezer tests the following results have been obtained. For the first part concerning the development of the cold front, Fig. 4, it is seen that the development of a well defined cold front diminishes with decreasing cylinder diameter.

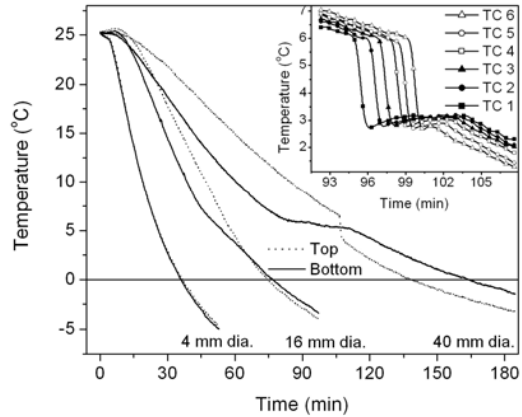

**FIG. 4.** Temperature versus time of a 4 mm, 16 mm, and 40 mm diameter cylinder with thermocouples placed at the top and bottom. **Inset:** Movement of cold front monitored by 6 thermocouples in a 60 mm diameter cylinder, 21 cm long, with thermocouples 2 cm from the top and spaced 1 cm apart. TC 1 was 14 cm from the bottom.

The second part involves the speed of the cold front as a function of diameter. As shown in Fig. 5, it is clear that the speed has a linear relationship with diameter. It is important to remember, as previously mentioned, that the development of the cold front diminishes with decreasing diameter thus making it difficult to measure its speed at small diameters.

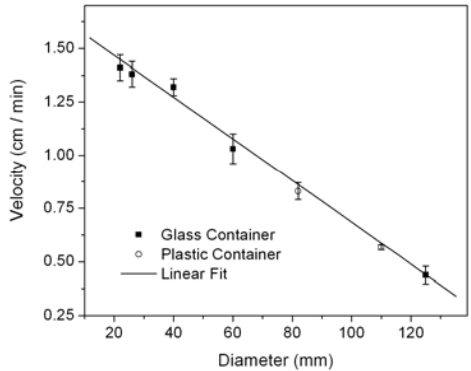

**FIG. 5.** Velocity versus diameter from the cooling of DI $H_2O$ in the freezer.

The velocities from the ice bath cooling were not calculated due to large experimental inconsistencies because of external induced turbulence.

Further experiments allowed us to determine how defined the cold front was by the slope of the temperature drop when the cold front reached the thermocouples, with the greater the slope relating to the more defined front. From that data we were also able to determine the shape of the cold front as seen in Fig. 6.

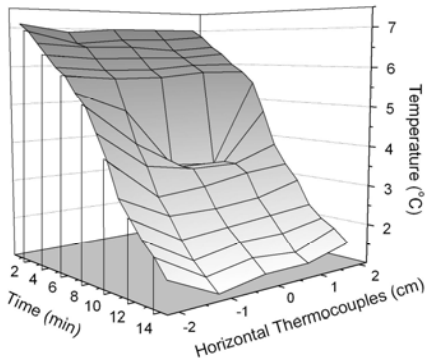

**FIG. 6.** A 3-Dimensional view of the response of 5 horizontal thermocouples located 4 cm from the top of a 40 mm diameter cylinder filled with water cooling from $\sim 20°C$ to $\sim 1°C$. A sudden change in temperature is registered when the cold front, rising from the bottom of the cylinder, arrived at the position of the thermocouples. Notice the temperatures between $7°C$ and $6°C$ were about equal for all thermocouples, as was true for the temperatures above $7°C$.

When cooling uniformly around the sides of the cylinder the cold front will be symmetrical and when there is differential cooling from the sides the cold front becomes asymmetrical.

Tests were also done with 99.9% $D_2O$ to see if a similar effect can be seen. These tests were not as extensive and were done only to show that the effect is still present. The tests were done with 2 identical containers in the freezer, one with DI $H_2O$ and the other with the $D_2O$. Both contained 3 thermocouples placed 0.5 cm from the top and spaced 2 cm apart in a 10 cm tall container. The results are shown in Fig. 7.

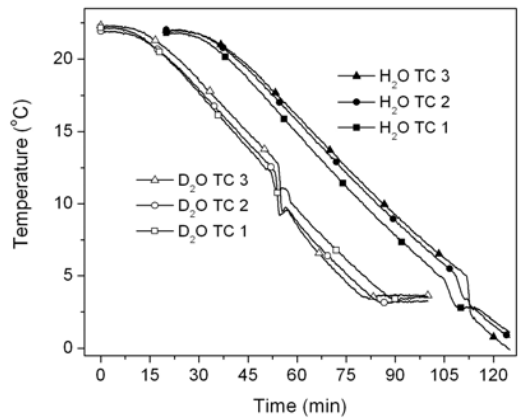

**FIG. 7.** Temperature versus time for $H_2O$ and $D_2O$ placed in identical containers. The starting times were offset by 20 minutes for graphing purposes. TC 1 is 4.5 cm from the bottom of a 10 cm container.

## DISCUSSION

As seen from Fig. 2 there is a distinct point at which each thermocouple registers a steep decrease in temperature. This is an indication of a cold front moving up the column of water. The cause for this cold front is the result of water having a maximum density of 0.99995 g/mL at $3.984°C$. As the water cools from say $20°C$, it becomes denser and thus starts to sink. This causes the water towards the bottom to cool faster because in addition to loosing heat from the side walls of the container, heat is being removed by denser colder water moving down from the top of the cylinder. As the water at the bottom cools to less than $\sim 4°C$, it starts to become less dense and therefore begins to rise. However, as it rises it comes in thermal contact with slightly warmer water causing it to heat up and become denser, thus sinking. This cycle continues and a cold front moves up the container until there is not enough water warmer than $\sim 4°C$ to keep the process going.

As the cold front begins to develop, the bottom thermocouple experiences a time of near constant temperature as compared to the 16 mm diameter cylinder. This can be seen in the bottom thermocouple of the 40 mm diameter cylinder in Fig. 4. The exact temperature depends on an equilibrium point between loss of heat through the outer wall of the cylinder and the addition of heat from the sinking water. The top thermocouple registers the arrival of the cold front with a sudden drop in temperature of a few degrees. It was determined that the smaller the diameter of the cylinder the less likely the cold front would develop. The 4 mm curve, in Fig. 4, shows no sign of the cold front, the 16 mm curve shows that the cold front develops and starts to move up the cylinder but does not make it to the top thermocouple, and the 40 mm curve shows the full development of the cold front. The reason for this is due to the fact that with the smaller diameter the heat is escaping through the sides faster than with a larger diameter. This faster transfer of heat through the sides does not allow the cold front time to develop. The inset to Fig. 4 shows a close up of how well defined the cold front becomes and the magnitude of the temperature drop. This setup consisted of 6 vertical thermocouples in a 60 mm diameter cylinder (20 cm long) with the thermocouples being 1 cm apart, starting 2 cm from the top.

We also show in Fig. 4 that the speed of the cold front is inversely proportional to the diameter of the cylinder. With cylinders of roughly the same height, the speed has a linear relationship with respect to diameter with it approaching zero at $170 \pm 5$ mm. We could not use a diameter smaller than 22 mm because the cold front becomes distorted at the top which makes it more difficult to calculate the speed. The speed does not appear to be dependent on the material of the container.

Knowing the vertical characteristics of the isotherm we set out to establish its horizontal profile. With 5 thermocouples placed 4 cm from the top we were able to determine how uniform it was and how defined it was when it rose to the positions of the thermocouples. The uniformity of the isotherm can be determined by calculating the time difference, between each thermocouple, of the arrival of the isotherm. The definition of the cold front is noted by the magnitude the temperature decreases when the thermocouples register the passing of the cold front. Figure 6 shows the result in 3D form and represents the decrease in temperature of each thermocouple versus time. Notice that each thermocouple reads the same temperature, as the water is cooled, prior to the arrival of the cold front and is displayed as the evenly spaced horizontal lines above the $6°C$. The cooling rate above $6°C$ was 0.27 $°C/\min$. When the cold front arrived, just after the 6 minute line in Fig. 6, the temperature of the center thermocouple dropped $2.57°C$ in 43.5 seconds (3.54 $°C/\min$). After the cold front passed, the water around the thermocouples resume cooling at approximately the same rate as before the cold front arrived (0.23 $°C/\min$). Although not clearly shown here, the cold front arrived at all 5 thermocouples at the same time. The larger temperature drop in the center shows that the cold front is more defined at the center than at the edge with a difference between them of 3.07 $°C/\min$. This uniformity is sensitive to equal cooling around the cylinder. The 12.5 cm diameter cylinder barely fit into the freezer and was not cooled evenly from all sides. Several unsuccessful attempts were made to correct for this and the symmetry improved but due to the lack of room in the freezer we were unable to obtain uniform cooling with a large container.

The rate that the water is cooled plays a major role in the development of the cold front. Tests show that if the column of water is cooled slowly enough, less than about 0.1 degrees per minute, the cold front may not develop. The reason for this maybe because when water is cooled at $< 0.1$ $°C/\min$ it cools over the entire volume at a nearly uniform rate and this does not allow the cold front to develop. In other words the entire volume of the water passes the density maximum at the same time.

The reason that the $D_2O$, Fig. 7, has the cold front develop at $\sim 11°C$ is because the maximum density of 1.1503 g/mL for $D_2O$ occurs at $11.2°C$.[16] Further tests are needed with the $D_2O$ to determine how closely its response is to that of $H_2O$.

## CONCLUSION

The study of water cooling while confined in a cylindrical column, with a diameter greater than ~20 mm, has shown that an isothermal line spontaneously develops and moves vertically upward as the temperature near the bottom approaches $\sim 4°C$. The development and movement of this isothermal line depends greatly on the diameter of the container used and less dependent on the height of the container. Although the cooling rate is a factor in the development of the cold front, it does not seem to be sensitive to small change, i.e. changing from glass to plastic as a container material showed no difference. The investigation of a possible unique difference between $H_2O$ and $D_2O$ as the temperature passes through the point of maximum density remains to be completed, see Fig. 7.

## AUTHOR CONTRIBUTIONS

a) Carried out experiments, analyzed data and wrote the manuscript; b) Carried out experiments and analyzed data; and c) Designed and carried out experiments and analyzed data.

## ACKNOWLEDGEMENT
We would like to thank S. M. Shafroth, M. Stephens, B. Poliks, T. Lowenstein and R. Sardo for their valuable conversations and help. We also thank Binghamton University Department of Physics, Applied Physics, and Astronomy for support.

## SUPPLEMENTAL MATERIAL

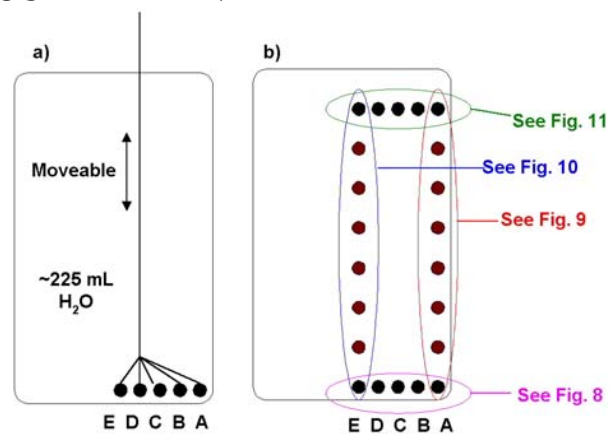

**FIG. 7.** This is the setup of an experiment that was conducted. Five thermocouples were placed on a moveable rod **(a)** and data was collected at 8 different positions **(b)**. Thermocouple A is 2 mm from the wall with B through E are 8, 14, 21, and 25 mm,

respectively. Position 0 is essentially directly on the bottom of the container and the thermocouples are moved up in increments of 1 cm to record the temperature as the water cools at that location. Each position represents an individual run where all 5 thermocouples were collecting temperature data during that run.

Refer to Figures 8-11 to see how the water cooled at the different thermocouple and position combinations. Figure 12 shows the difference in cooling between Coke, Diet Coke, and Water.

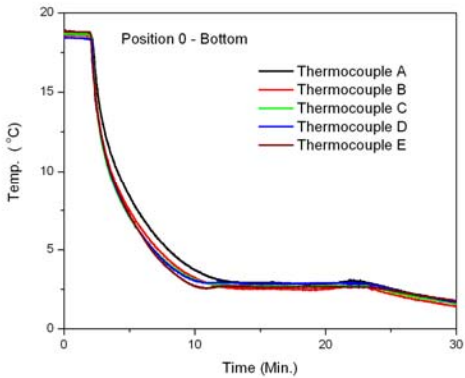

**FIG. 8.** Temperature versus time when the thermocouples were at Position 0.

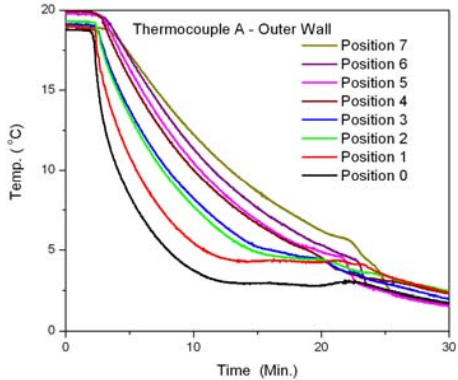

**FIG. 9.** Temperature versus time for thermocouple A as it is moved up the container of water. Again each position represents an individual run meaning that data was collected at Position 0, the water was allowed to warm, then the thermocouples were moved to Position 1 and the data was collected again as the water was cooled.

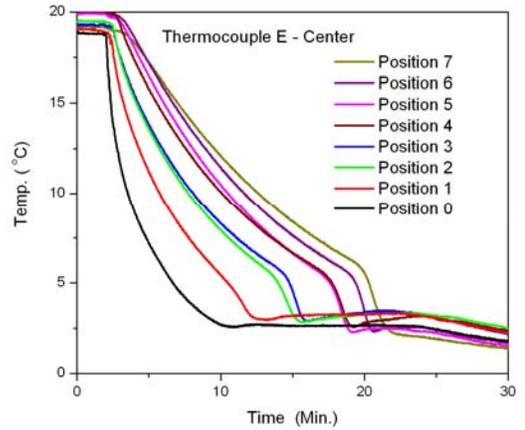

**FIG. 10.** Temperature versus time for thermocouple E as it was moved up the container of water.

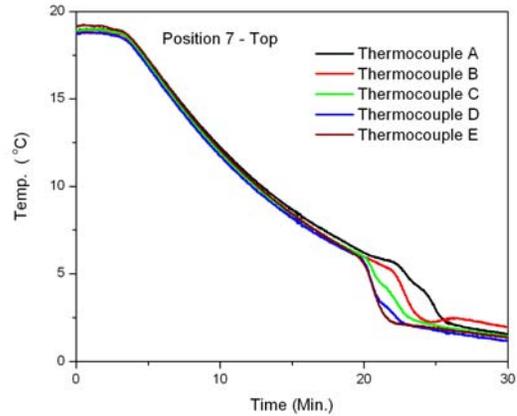

**FIG. 11.** Temperature versus time when the thermocouples were at Position 7.

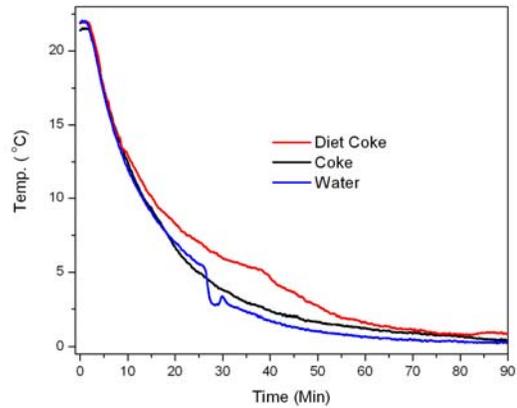

Fig. 12. Temperature versus time for the cooling curve with one thermocouple placed in freshly opened

20 oz containers of Coke, Diet Coke, and Tap Water. The containers were placed in an ice bath to cool. (Coke: 65 grams Sugar, 75 mg Na; Diet Coke: 0 grams Sugar, 75 mg Na)